\documentclass[aps,pre
,preprint
,showpacs
,groupedaddress
,superscriptaddress
,nobibnotes
]
{revtex4-1}
\usepackage{amsmath,amsthm,amssymb}
\usepackage{array}
\usepackage{subfigure}
\usepackage{graphicx}
\usepackage{fancyhdr}
\usepackage{color}
\usepackage{extarrows}
\usepackage{enumerate}
\usepackage{epstopdf}
\usepackage{bm}
\usepackage[colorlinks,
          linkcolor=black,
            citecolor=black,
            urlcolor=blue
           ]{hyperref}
\usepackage{natbib}
\usepackage{comment}

\renewcommand{\vec}[1]{\mathbf{#1}}

\usepackage{xr}
\usepackage{tabularx}

\usepackage[normalem]{ulem}

\begin{document}

\title{Stochastic Gradient Descent and Anomaly of Variance-flatness Relation in Artificial Neural Networks}
\author{Xia Xiong}
\affiliation{Shanghai Center for Quantitative Life Sciences \& Physics Department, Shanghai University, Shanghai 200444, China}
\author{Yong-Cong Chen}
\email[Corresponding author: ]{chenyongcong@shu.edu.cn}

\affiliation{Shanghai Center for Quantitative Life Sciences \& Physics Department, Shanghai University, Shanghai 200444, China}
\author{Chunxiao Shi}
\affiliation{Shanghai Center for Quantitative Life Sciences \& Physics Department, Shanghai University, Shanghai 200444, China}
\author{Ping Ao}
\affiliation{Colloge of Biomedical Engineering, Sichuan University, Chengdu 610065, China}
\begin{abstract}
Stochastic gradient descent (SGD), a widely used algorithm in deep-learning neural networks has attracted continuing studies for the theoretical principles behind its success. A recent work reports an anomaly (inverse) relation between the variance of neural weights  and the landscape flatness of the loss function driven under SGD [Feng \& Tu, PNAS 118, 0027 (2021)]. To investigate this seemingly violation of statistical physics principle, the properties of SGD near fixed points are analysed via a dynamic decomposition method. Our approach recovers the true ``energy'' function under which the universal Boltzmann distribution holds. It differs from the cost function in general and resolves the paradox raised by the the anomaly. The study bridges the gap between the classical statistical mechanics and the emerging discipline of artificial intelligence, with potential for better algorithms to the latter.

\end{abstract}

\maketitle

\renewcommand{\vec}[1]{\mathbf{#1}}
\newcommand{\bvec}[1]{\mbox{\boldmath $#1$}}
\newcommand{\cmmnt}[1]{\ignorespaces}

\section{Introduction}

An artificial neural network (ANN) is a machine learning platform which resembles a human neural system and its learning process. The system includes at least three layers: an input layer, a hidden layer and an output layer. When there are multiple hidden layers, it is called a deep neural network (DNN), whose efficient training has been a major topic in the emerging  discipline \cite{lecun2015deep,goodfellow2016deep,aggarwal2018neural}. Most DNNs engage routines that adjust the weights of neural connections and minimize the so-called loss function for the learning process \cite{le2011optimization,martens2010deep,young2015optimizing}. Among them, stochastic gradient descent (SGD) has been a particularly successful algorithm which appears in almost all DNN applications. But the theoretical principle behind its tremendous success is still an on-going research topic \cite{advani2013statistical,baldassi2016unreasonable,zhang2021understanding}. Possibly, the high degree of stochastic anisotropy associated with SGD enables the optimization to escape from local minima en route to the best configuration \cite{chaudhari2018stochastic}.

In search for the answer, there has been a number of ideas borrowed from the classical statistical physics \cite{zhang2018energy,feng2021phases,carleo2019machine,mehta2019high}. Recently Feng et al \cite{Fenge2015617118} regarded the loss function as resembling the ``thermodynamic'' energy function for a ``physical'' system. They derived an equivalent stochastic equation and  performed principal component analysis (PCA) on the weight dynamics. The latter reduced the dimensions of the complex stochastic process in accordance with their significance: The weights were projected onto the PCA axes and further analyzed. When the loss function \cite{ghorbani2019investigation,li2018visualizing} was characterized by its distribution width on the PCA directions, they found a robust anomaly between the weight variance and the flatness of the landscape, in striking contrast to what would follow from the conventional Boltzmann distribution. Apparently, ANN appeared to violate a fundamental principle of physics and display what they referred to as an ``inverse Einstein relation''.

This work investigates such unconventional characteristic in detail. We start with a formalism of stochastic decomposition developed by Ao et al. \cite{ao2004potential}. The method obtains for the dynamics a Lyapunov functional which is an alternative to the cost function. It is a close analogue to the energy function of a physical system under which statistical principle holds\cite{Kwon13029}. For a linear system, the deterministic driving force in the process can be strictly decomposed into two parts. One of them leads to balance of probabilistic distribution and the other to cyclic motion on the surface of constant potential function. Together they offer near a fixed point a Boltzmann-like probability distribution as well as flux-carrying stationary states without detailed balance. This methodology has been applied to a number of practical systems \cite{chen2020global,shi2023computation,PNSwork}, in particular, complex networks for biological studies \cite{yuan2017cancer}.

In our approach, the covariance matrix for the dynamics is in fact inversely proportional to the energy matrix $\textbf{U}$ for the stochastic potential. Such a relationship well resolves the paradox raised by the anomaly in the variance-flatness relation (VFR). We further examine the scaling behaves of the weight variance and the flatness by explicit reconstruction of the cost function in the linear region from the fluctuation data (i.e. diffusion matrix). The study of stochastic decomposition in ANN may offer better algorithm that can address problems such as the well-known ``forgetting'' catastrophe \cite{robins1995Catastrophic,kirkpatrick2017overcoming} while performing multi tasks.

The work is organized as follows. In the next section we clarify the VFR anomaly in SGD-based ANNs reported by Feng et al. In Sect.~\ref{III} we first briefly review the stochastic decomposition and identify a direct relationship between the covariance matrix and the stochastic potential. This allows us to look further in Sect.~\ref{IV} into the question of VFR with respect to the Boltzmann distribution under proper energy function. Later in Sect. \ref{V} we re-examine the scaling behavior of VFR under the new approach, with an outlook to the algorithm optimization in ANN, followed by a brief summary and discussion in Sect.~\ref{VI}. 

\section{The Continuous limit of SGD}
\label{II}

ANN optimizes its state vector, i.e. weights and biases through a training process. In general, the loss function can fairly complicated with a huge parameter space \cite{bray2007statistics,beer2006parameter,amari1996neural}. On a gradient descent scheme, the state vector advances a certain step from the current position along the opposite direction of the gradient. The latter is re-calculated at the new position to continue the optimization. When the data used for training are randomly selected mini-batches, the scheme is known as the stochastic gradient descent  \cite{rattray1998natural,sohl2014fast}, or simply SGD. Specifically, the state vector $\bm{\omega}_{k}$, an array of weights in question at the $k^{\text{th}}$ stage is now updated via
\begin{equation}
\label{SGDupdate}
\bm{\omega}_{k+1} = \bm{\omega}_{k} - \alpha\nabla_{\omega} L^{\mu}(\bm{\omega}_{k}),
\end{equation}
where $\nabla_{\omega} L^{\mu}(\bm{\omega})$ is the gradient over the the loss function of the $\mu^{\rm{th}}$ mini-batch (of size $B\gg 1$) and $\alpha$ is a small learning rate. Note that throughout this work, a boldface symbol is used for matrix or vector, while the same in normal face is for the underlying matrix element(s).

Following Feng et al \cite{Fenge2015617118}, the so-called continuous-time limit \cite{sompolinsky1988chaos} of SGD can be cast into a conventional set of stochastic differential equation as
\begin{equation}
\label{DNN_SDE}
\dot{\bm{\omega}} = -\nabla_{{\omega}}L(\bm{\omega}) + \bm{\eta}(t).
\end{equation}
In Eq.~(\ref{DNN_SDE}) we define $\dot{\bm{\omega}} \equiv \text{d}\bm{\omega}/\text{d}t$ with $\text{d}t\equiv \alpha$, and $L(\bm{\omega})$ is the ``correct'' loss function for the learning process (an average over all samples, cf. below). The SGD noise term  $\bm{\eta} \equiv -\nabla_{{\omega}}(L^{\mu}-L)$  arises from the variation between a mini-batch and the full-batch loss functions. The noise has zero ensemble average $\langle\bm{\eta}\rangle_{\mu}=0$, but non-zero variance $\langle\eta_{i}(t)\eta_{j}(t')\rangle_{\mu}$ for $t=t'$.

The learning set is usually assumed to be sufficiently large so that there is little correlation between the mini-batches.
In a SGD process when a mini-batch is sampled with replacement, the variance reads\cite{chaudhari2018stochastic}
\begin{equation}
\label{minibatchGradientVariance}
\left\langle(\nabla L^{\mu} - \nabla L)(\nabla L^{\mu} - \nabla L)\right\rangle_{\mu}
= \vec{D}(\bm{\omega})/B.
\end{equation}
Here $\vec{D}(\bm{\omega})$ is a diffusion matrix independent of the mini-batches, which can be computed from
\begin{equation}
\label{D_def}
\vec{D}(\bm{\omega})\approx\left(\frac{1}{N_{L}}\sum^{N_{L}}_{k=1}\nabla L_{k}\nabla L_{k}\right)-\nabla L(\bm{\omega})\nabla L(\bm{\omega}),
\end{equation}
where $L_{k} \equiv L_{k}(\bm{\omega})$ is the loss function for the $k^{\text{th}}$ sample,  $L(\bm{\omega})\equiv (1/N_{L})\sum_{k=1}^{N_{L}}L_{k}$, and $N_{L}$ is the total size of the learning set. A slight variation but essentially the same for $B/N_L\ll 1$  can be obtained for sampling without replacement.  Evidently, to cast $\langle \eta_{i} (t)\eta_{j}(t')\rangle_{\mu}$ into the form $2 \epsilon D_{ij}\delta (t-t')$ of Eq.~(\ref{SDE}) below] the noise strength $\epsilon$ in Eq.~(\ref{Stationary_distribution}) reads $\alpha/2B$.

\subsection{Anomaly of Variance-flatness Relation}

We next summarize the main result obtained by Feng et al \cite{Fenge2015617118}. After carrying out PCA on the SGD process, the weight dynamics can be projected out for variations in the principal axes, $\bm{\omega}(t) = \langle\bm{\omega}\rangle_{T} + \sum_{i}\theta_{i}(t)\vec{p}_{i}$. Here $\langle\bm{\omega}\rangle_{T}\equiv \bm{\omega_{0}}$ is the average weight vector in a particular epoch time of length $T$, $\vec{p}_{i}$ is the $i^{\rm{th}}$ principal base vector and $\theta_{i}(t)$ is the projection on the direction. Let the loss function profile along $\bm{p}_{i}$ be $L_{i}(\delta\theta)\equiv L(\bm{\omega_{0}}+\delta\theta\bm{p}_{i})$, numerous data simulations found that $L_{i}$ becomes flatter with the increase of $i$.

To quantify the behavior, a flatness parameter can be introduced as below \cite{hochreiter1997flat,chaudhari2019entropy,baldassi2020shaping}. $F_{i} = \theta^{r}_{i} - \theta^{l}_{i}$ ($\theta^{l}_{i} < 0, \theta^{r}_{i} > 0$) is taken as the difference between the nearest two points of equal value near the minimum found by SGD. Specifically $L_{i}(\theta^{l}_{i}) = L_{i}(\theta^{r}_{i}) = e\times L_{0}$, where $e$ is the Euler's number and $L_{0}$ is the
loss function at local minimum. Now $F_{i}$ is found to increase with PCA index $i$, whereas the SGD variance $\sigma^{2}_{i}$ decreases with $i$. Finally the VFR follows approximately a power-law behavior,
\begin{equation}
\label{Variance_flatness}
\sigma^{2}_{i}\sim F^{-4}_{i}.
\end{equation}

The above result is counter-intuitive for the flatter the ``energy function'' the less diffusive the dynamics.
In statistical physics where the loss function $L$ assumed the role of the energy function, the equilibrium probability distribution of $\bm{\theta}$ would follow Boltzmann distribution, i.e. $P(\bm{\theta}) = \exp[-L(\bm{\theta})/T]$ (with ``empirical temperature'' $T$). One would then have $\sigma^{2}_{i} \propto F^{2}_{i}$ so that $\sigma^{2}_{i}/F^{2}_{i}\sim$ constant, in stark contrast to Eq.~(\ref{Variance_flatness}). The latter apparently would violate a fundamental law of physics.

\section{Covariance matrix of the stochastic dynamics}
\label{III}

As the covariance of the stochastic process plays a pivot role in the PCA \cite{abdi2010principal}, we next introduce an alternative method of stochastic decomposition to analyze the dynamics. In doing so we gain the knowledge of a crucial relationship between the covariance and the ``true'' energy function that ought be used for the Boltzmann distribution, fulfilling a major goal of the present work.

\subsection{A Review of the Stochastic Decomposition}

A large class of stochastic processes in nature can be modeled by stochastic differential equations (SDE) \cite{van1992stochastic} in a generic  form
\begin{equation}
\label{SDE}
\dot{\textbf{x}}=f(\textbf{x})+\bm{\zeta}(t),
\end{equation}
where $\textbf{x}$ stands for a set of dynamical variables, a general state vector of a $N$-dimensional system. Here $f(\vec{x})$ and $\bm{\zeta}(t)$ are respectively the deterministic and a Markovian driving force for the dynamics. In many cases $\bm{\zeta}(t)$ can be represented by a (functional of) Gaussian white noise with $\langle\bm{\zeta}(t)\rangle=0$ and semi-positive definite variance $\langle\bm{\zeta}(t)\bm{\zeta^}{\tau}(t')\rangle=2\epsilon\vec{D}\delta(t-t')$. The superscript $\tau$ denotes the transpose of the underlying vector/matrix, $\langle\cdots\rangle$ is the average over the Gaussian distribution, and $\delta(t)$ is the Kronecker delta function for the Markovian process. Evidently $\vec{D}$ is a diffusion matrix with $\epsilon$ being the noise strength, which plays the role of temperature of the stochasticity. For the analysis below, we will focus on dynamics near a fixed point so that the first term in Eq.~(\ref{SDE}) may be approximated by $f_{i}(x)=-F_{ij}x_{j}$.  Namely, it reduces to
\begin{eqnarray}
\label{Linear_approximation}
 \dot{\vec{x}}=-\textbf{F}\,\textbf{x}+\bm{\zeta}(t).
\end{eqnarray}

The aforementioned stochastic decomposition seeks to re-cast the dynamics into the following ``canonical'' form,
\begin{eqnarray}
\label{Decomposed_SDE}
 \left(\textbf{S}+\textbf{A}\right)\dot{\textbf{x}} = -\textbf{U}\,\textbf{x} + \bm{\xi}(t).
\end{eqnarray}
Among the terms, $\textbf{U}$ is a symmetric potential matrix which gives rise to a stochastic potential or an energy function for the whole system
\begin{eqnarray}
\label{Stochastic_Potential}
u(\textbf{x})= \textbf{x}^{\tau}\textbf{U}\,\textbf{x}/2.
\end{eqnarray}
$\textbf{S}$ is symmetric, semi-positive definite. It represents the dissipative dynamics which, in the absence of fluctuations, causes monotonic decrease of $u(\textbf{x})$. The canonical noise $\bm{\xi}(t)$ is associated with $\textbf{S}$ by a generalized fluctuation-dissipation theorem $\left\langle\bm{\xi}(t)\bm{\xi}^{\tau}(t')\right\rangle = 2\epsilon \textbf{S}\delta(t-t')$. On the other hand, $\textbf{A}$ is antisymmetric and conserves $u(\textbf{x})$. With the canonical form it can be shown \cite{ao2004potential} that the Boltzmann-like distribution
\begin{equation}
\label{Stationary_distribution}
\rho(\textbf{x})\propto \exp\{-u(\textbf{x})/\epsilon\}
\end{equation}
is a stationary distribution for the system. That is, the potential function so constructed bears the essence of energy in statistical physics.

Going back to Eq.~(\ref{Linear_approximation}), the two norms of noise are related by
\begin{eqnarray}
\label{Noise}
 \bm{\xi}(t) = \left(\textbf{S}+\textbf{A}\right)\bm{\zeta}(t),
\end{eqnarray}
\begin{eqnarray}
\label{S_to_D}
\textbf{S} = (\textbf{S}+\textbf{A})\textbf{D}(\textbf{S}-\textbf{A}).
\end{eqnarray}
Furthermore, take the inverse of $(\textbf{S}+\textbf{A})^{-1}$ and break it down to symmetric $\tilde{\textbf{D}}$ and anti-symmetric $\textbf{Q}$. We find $(\textbf{S}+\textbf{A})\textbf{Q}(\textbf{S}-\textbf{A}) = -\textbf{A}$, and $\tilde{\textbf{D}}\equiv \vec{D}$ [as both $\tilde{\vec{D}}$ and $\vec{D}$ satisfy the same  Eq.~(\ref{S_to_D})].

To obtain the potential function, we rewrite the force matrix $\textbf{F}$ as
\begin{equation}
\label{Force_matrix}
\textbf{F} = (\textbf{S}+\textbf{A})^{-1}\textbf{U} = (\textbf{D}+\textbf{Q})\textbf{U}.
\end{equation}
Using the symmetry of $\textbf{U}$, $\textbf{D}$ and $\textbf{Q}$, $\textbf{U}$ and $\textbf{Q}$ can be determined uniquely from $\textbf{F}$ and $\textbf{D}$ via
\begin{equation}
\label{F_U}
\textbf{F}\textbf{U}^{-1}+\textbf{U}^{-1}\textbf{F}^{\tau}=2\textbf{D},
\end{equation}
\begin{equation}
\label{F_D_U}
\textbf{F}\textbf{Q}+\textbf{Q}\textbf{F}^{\tau}=\textbf{F}\textbf{D}-\textbf{D}\textbf{F}^{\tau}.
\end{equation}
The above results hold near a stable fixed point where real parts of eigenvalues of $\textbf{F}$ are positive, a condition that is assumed to be satisfied in what follows. Further details can be found in e.g. \cite{Kwon13029}.

\subsection{Evaluation of Covariance Matrix}

To proceed further we solve Eq.~(\ref{Linear_approximation}) for the system as,
\begin{equation}
\label{Exact_solution }
\textbf{x}(t)=e^{-\textbf{F}t}\left[\textbf{x}(0)+\int^{t}_{0}e^{\textbf{F}t'}\bm{\zeta}(t')\text{d}t'\right].
\end{equation}
The covariance is calculated from regulated $\textbf{x}$ with zero means. Namely, $\tilde{\textbf{x}}(t)=\textbf{x}(t)-(1/t)\int^{t}_{0}\textbf{x}(t')\text{d}t'$. Near a stable fixed point the homogeneous part decays exponentially. Therefore at $t\rightarrow \infty$ the covariance matrix $\bm{\Sigma}$ can be readily obtained as
\begin{equation}
\begin{aligned}
\label{Covariance_matrix}
\bm{\Sigma}&=\left\langle\tilde{\textbf{x}}(t)\tilde{\textbf{x}}^{\tau}(t)\right\rangle_{t\rightarrow\infty}
&=2\epsilon\int_{-\infty}^{0}\text{d}t\,e^{\textbf{F}t}\textbf{D}e^{\textbf{F}^{\tau}t}.
\end{aligned}
\end{equation}
The integration can be further carried out \cite{han2021fluctuation} to get
\begin{equation}
\begin{aligned}
\label{CovarianceMatrixWithForceMatrix}
\textbf{F}\bm{\Sigma} + \bm{\Sigma}\textbf{F}^{\tau} &=2\epsilon\int_{-\infty}^{0}\text{d}t\,\frac{\text{d}}{\text{d}t}\left[e^{\textbf{F}t}\textbf{D}e^{\textbf{F}^{\tau}t}\right]
&=2\epsilon\textbf{D}.
\end{aligned}
\end{equation}

It follows from Eq.~(\ref{F_U}) that the covariance matrix and the potential energy matrix are the mutual inverse of each other,
$\bm{\Sigma}=\epsilon\textbf{U}^{-1}$.
The largest principal components of $\bm{\Sigma}$ correspond to the eigenstates of $\textbf{U}$ with smallest eigenvalues.


\section{Statistical and Scaling Properties of SGD}

Returning to Eq.~(\ref{DNN_SDE}), it is evident that we can set $\bm{\omega}\equiv \vec{x}$ and map the SGD noice $\bm{\eta}(t)$ to $\bm{\zeta}(t)$ in the current context, hence establish a relationship between the loss function, characterized by the matrix $\vec{F}$ via $L(\textbf{x})= L_{0} + \textbf{x}^{\tau}\textbf{F}\,\textbf{x}/2$ near a fixed point,  and the stochastic potential set by $\vec{U}$, cf. Eq.~(\ref{Stochastic_Potential}). It suggests that the mystery of anomaly VFR  may be traced back to the choice of the energy function. For an ANN, if the loss function was incorrectly assumed as the energy function, the relationship between variance and flatness would not conform to the usual statistical physics. Instead, they could show the counter intuitive VFR characteristic as seen in the last section.

\subsection{Boltzmann Distribution of Neural Weights}
\label{IV}

The correct Boltzmann distribution, i.e. Eq.~(\ref{Stationary_distribution}), follows when the stochastic potential is recognized as the proper energy function.  Since $\bm{\Sigma}\sim \textbf{U}^{-1} $, the value of variance increases as the eigenvalue of $\textbf{U}$ is reduced. Hence the landscape of $u(\vec{x})$ from Eq.~(\ref{Stochastic_Potential}) becomes flatter, leading to larger flatness $F_{i}$ in terms of the new energy function. This is a positive correlation of VFR, contrary to what appears in Eq.~(\ref{Variance_flatness}). It resolves naturally the conceptual difficulty raised by Feng et al \cite{Fenge2015617118} as outlined in Sect.~\ref{II}.

\subsection{Data Simulations and Scaling Properties}
\label{V}
The explicit relationship between $\vec{F}$, $\vec{D}$, and $\bm{\Sigma}$ in Eq.~(\ref{CovarianceMatrixWithForceMatrix}) allows us to recover the details of the loss function through the $\vec{F}$ matrix near a fixed point. It in turn can be used to re-examine the scaling properties reported in by Feng et al \cite{Fenge2015617118}. For that purpose additional data simulations are carried out in the supplementary information (SI) \cite{supplementary}. The main results are reported below.

\begin{figure*}[!ht]
\begin{tabular}{ccc}
\includegraphics[width=0.4\textwidth]{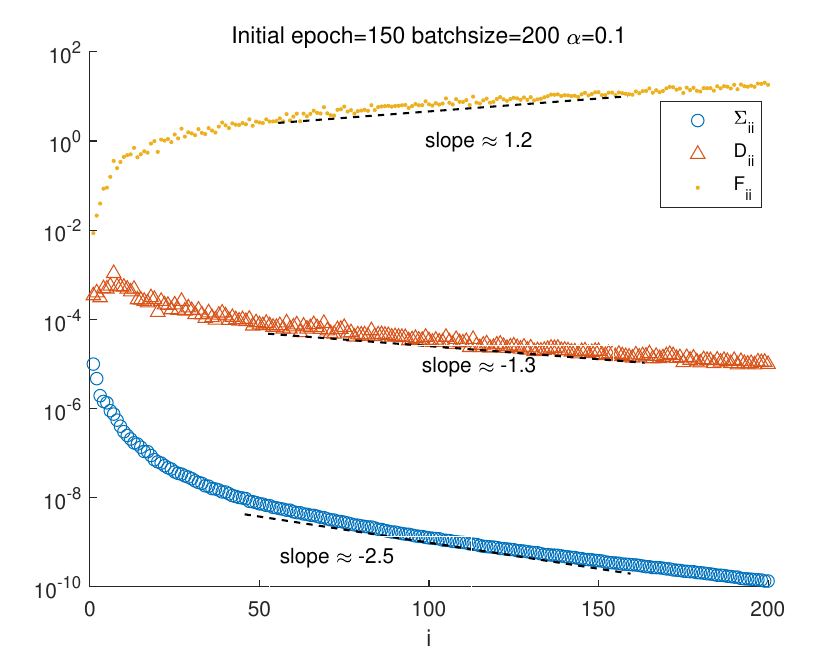} &
\includegraphics[width=0.4\textwidth]{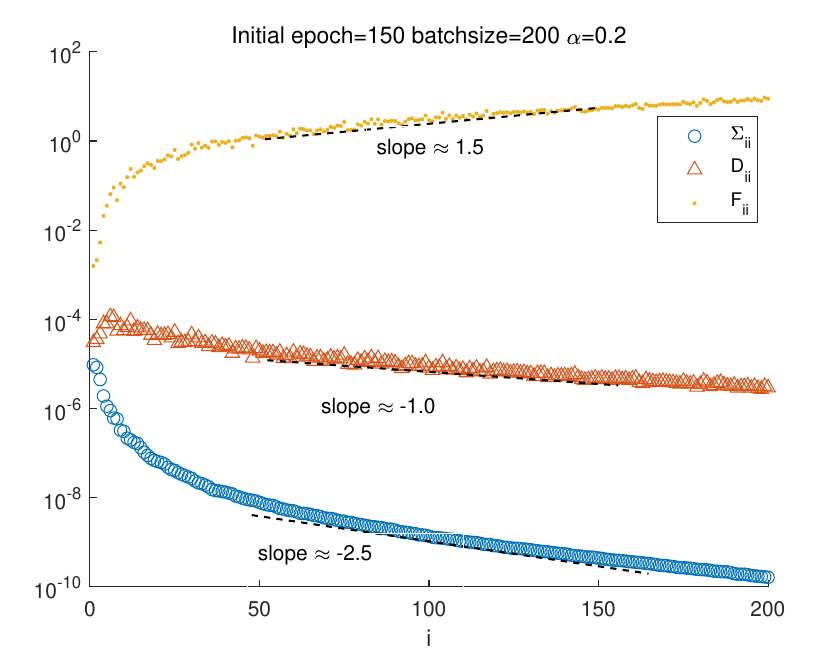} \\
(a) & (b) \\
\end{tabular}
\protect\caption{The diagonal matrix elements of $\bm{\Sigma}$, $\vec{D}$ and $\vec{F}$ at different learning rates under the eigen-space of $\bm{\Sigma}$. The horizontal axis $i$ represents the PCA index.  (a)Learning rate $\alpha=0.1$, which shows the scaling behavior as $\Sigma_{ii}\propto i^{-2.5}$, $D_{ii}\propto i^{-1.3}$, and $F_{ii}\propto i^{1.2}$. (b) $\alpha=0.2$, with slightly different scaling indexes as $\Sigma_{ii}\propto i^{-2.5}$, $D_{ii}\propto i^{-1.0}$, $F_{ii}\propto i^{1.5}$.
}
\label{FDCmatrixForCpre}
\end{figure*}

In an exemplary set up, we explore a system of fully connected neural network with two hidden layers. The weight elements between the two hidden layers is extracted to form the set of weights $\bm{\omega}$, i.e. the state vector $\vec{x}$ in the current context. The covariance matrix of the weights is calculated and diagonalized with eigenvalues ranked in descending order. The eigenvectors can be used as the convenience basis for the weight space.  Furthermore, the diffusion matrix $\vec{D}$ can be sampled directly via Eq.~(\ref{minibatchGradientVariance}) or calculated via Eq.~(\ref{D_def}). We found that the difference is negligible. Note that $\vec{F}$ (being second derivatives on the loss function) is also symmetric hence it can be easily obtained via Eq.~(\ref{CovarianceMatrixWithForceMatrix}) in the diagonal representation of $\bm{\Sigma}$, cf. SI \cite{supplementary} for more details. The diagonal matric elements of $\vec{D}$ and $\vec{F}$, $D_{ii}$ and $F_{ii}$ are shown in Fig.~\ref{FDCmatrixForCpre}.

There are a few key points taken away from the results. First of all, $D_{ii}$ of the diffusion matrix $\vec{D}$ also decreases with the PCA index $i$, {\em albeit} at a slower rate than $\Sigma_{ii}$. Secondly, it is (learning rate) $\alpha$-dependent, which appears to suggest that larger $\alpha$'s lead to local minima of loss function that are smoother and less fluctuating under SGD. In the vicinity of a local minimum, we have $F_{ii}\sim F^{-2}_{i}$ by definition of flatness in the loss function. Again were that function taken as the energy function for statistical distribution, we would have the same anomaly that $\Sigma_{ii}\times F_{ii}\sim \Sigma_{ii}/F^{2}_{i}$ does not scale to a constant, rather it diminishes as $i$ increases.  On the other hand, we do not find the inverse relationship of Eq.~(\ref{Variance_flatness}) either. The discrepancy may be due to the flatness definition employed by Feng et al \cite{Fenge2015617118}. Their definition is more an intuitive and global one and the measure is obtained rather far away from the local minimum. Another factor might arise from the limitation on the continuous-time approximation of SGD. More remains to be explored.

\subsection{Potential Algorithm Improvement}

The stochastic decomposition itself can reveal rich structural features. In a related study \cite{PNSwork}, the eigenvalues and eigenvectors of the $\textbf{S}$ and $\textbf{A}$ matrices [cf. Eq.~(\ref{Decomposed_SDE})] are analyzed to a great extent. Under an isotropic diffusion environment, the dissipation matrix $\vec{S}$ can be expressed as the direct sum of multiple matrices, $\vec{S}=\vec{s}_{1}\bigoplus\vec{s}_{2}\bigoplus...\vec{s}_{N/2}$, where $\vec{s}_{i}=s_{i}\vec{1}$ with the $2\times2$ unit matrix $\vec{1}$. Here, $s_{i}>0$ is the $i^{th}$ eigenvalue which can be sorted by $s_{1}\leq s_{2}\leq s_{3} \leq...\leq s_{N/2}$. In the same subspaces $\vec{A}$ can also be diagonalized into a direct sum of pairs of $2\times2$ antisymmetric matrices. The system is often found circulating in the subspaces with small $s_{i}$'s, i.e. it shows vortex-like behavior when the damping or viscosity is small.

In the current context, $\vec{S}$, $\vec{A}$ can be obtained via Eq.~(\ref{Force_matrix}) whereas $\vec{Q}$ can be obtained from Eq.~(\ref{F_D_U}) once $\vec{F}$ is known. The presence of $\vec{Q}$ is an indication that $\vec{D}$ and $\vec{F}$, as in the case of anisotropic $\vec{D}$, do not commute. The magnitude of $\vec{Q}$ may appear insignificant when compared to $\vec{F}$. But the correct comparison should be drawn between $\vec{F}$ and $\vec{D}$ as they are the odd and even parts of one matrix in Eq.~(\ref{Force_matrix}). After we-scaling the state variables so that $\vec{D}$ is transformed into the isotropic identity matrix $\vec{I}$, $\vec{Q}$ is enhanced by several order-of-magnitude. A large $\vec{Q}$ will lead to small $\vec{S}$. Hence there will be abundance of vortices to emerge. More details concerning $\vec{Q}$ are presented in the SI \cite{supplementary}.

The characteristic may be useful for a solution on the well-known catastrophic weight-forgetting difficulty. The latter refers to the problem that when a neural network learns a new task under the premise that it has already learned an old one, it often performs poorly on the first task afterwards. This can be understood as due to heavy shifting of neural weights in learning the new task such that the old settings are forgotten \cite{robins1995Catastrophic}. The structural vortices identified above may be used to confine the drifting  of weights in a more lenient way than a brute-force pin-down of them to around the first task. Namely, the weights are allowed to circulate in the weight space instead.  Multiple tasks can be associated to vortices from different hidden layers. Any success will be an indication that our approach may indeed be adopted in ANNs to benefit future development.

\section{Discussion}
\label{VI}

To summarize, a dynamic decomposition is used to analyze the stochastic properties near fixed points of  SGD in ANNs. Controversies from previous studies regarding some fundamental statistical principle such as  the Einstein relation are resolved by the identification of stochastic potential function, i.e. the proper energy function which differs from the loss function used in ANNs when the stochasticity of SGD is concerned. In addition, the method and data simulations have revealed further characteristics and structures of SGD.

The success of stochastic dynamic decomposition in ANNs brings us insights into the theoretical understanding of deep learning systems, which may offer ideas for the subsequent development of better algorithms in artificial intelligence. For example, we are able to identify the non-vanishing transverse $\vec{Q}$ matrix for the stochastic process. The latter are found closely related to the vortex-like circulations in similar systems \cite{PNSwork}. Such features may be further explored in the current field. It may be used to construct a more natural strategy targeting the catastrophic forgetting problem in DNNs.

\section{acknowledgments}
This work was supported in part by the Natural Science Foundation of China No. 16Z103060007 (PA). One of us (XX) thanks Dr. Pratik Chaudhari for  communications and free codes from his work. Thanks to SHANGHAI NANOBUBBLE TECHNOLOGY CO.,LTD. for supporting this work.

\appendix
\renewcommand{\thefigure}{S\arabic{figure}}
\setcounter{figure}{0}

%

\end{document}


\title{Supplementary Information for Stochastic Gradient Descent and Anomaly of Variance-flatness Relation in Artificial Neural Networks}
\author{Xia Xiong}
\affiliation{Shanghai Center for Quantitative Life Sciences \& Physics Department, Shanghai University, Shanghai 200444, China}

\author{Yong-Cong Chen}
\email{chenyongcong@shu.edu.cn}
\affiliation{Shanghai Center for Quantitative Life Sciences \& Physics Department, Shanghai University, Shanghai 200444, China}
\author{Chunxiao Shi}
\affiliation{Shanghai Center for Quantitative Life Sciences \& Physics Department, Shanghai University, Shanghai 200444, China}
\author{Ping Ao}
\affiliation{Colloge of Biomedical Engineering, Sichuan University, Chengdu 610065, China}

\begin{abstract}
This supplementary information (SI) covers additional computational and data simulations used or cited in the main article \cite{mainwork}. Specific details related to the construction of the neural networks and the composition of the data are presented, along with the extraction methods from the networks.  Then  matrices discussed in the main work under stochastic decomposition are evaluated. Results are graphically illustrated for ease of understanding.
\end{abstract}
\maketitle


\renewcommand{\vec}[1]{\mathbf{#1}}
\newcommand{\bvec}[1]{\mbox{\boldmath $#1$}}

\renewcommand{\theequation}{S\arabic{equation}}
\renewcommand{\thefigure}{S\arabic{figure}}
\renewcommand{\thesection}{\Alph{section}}
\renewcommand{\thesubsection}{\Alph{section}.\arabic{subsection}}
\newcommand{\cmmnt}[1]{}

\renewcommand{\theequation}{S\arabic{equation}}
\renewcommand{\thefigure}{S\arabic{figure}}
\renewcommand{\thetable}{S\arabic{table}}

\tableofcontents
\newpage

\section{Data Composition and Network Structure}
\label{SI}

In this supplemental work, we conduct exemplary simulations of artificial neural network (ANN), extracting and analyzing the dynamics of neural weights driven under stochastic gradient descent (SGD). A set of fully-connected neural networks are constructed with the MNIST dataset of handwritten digits
commonly employed in deep learning work simulations \cite{lecun1998mnist}. The dataset of handwritten digital photographs has four parts, consisting of $60,000$ training data, $6,0000$ training labels,  $1,0000$ test data, and $10,000$ test labels. The networks have multiple structures, with the layers connected to each other by the Relu function. No bias is set in each network layer for the convenience of computation and analysis. The usual cross-entropy loss is taken as the loss function. Networks with different learning rates, different numbers of layers, and different batchsizes are simulated.

\begin{figure*}[!ht]
\begin{tabular}{ccc}
\includegraphics[width=0.4\textwidth]{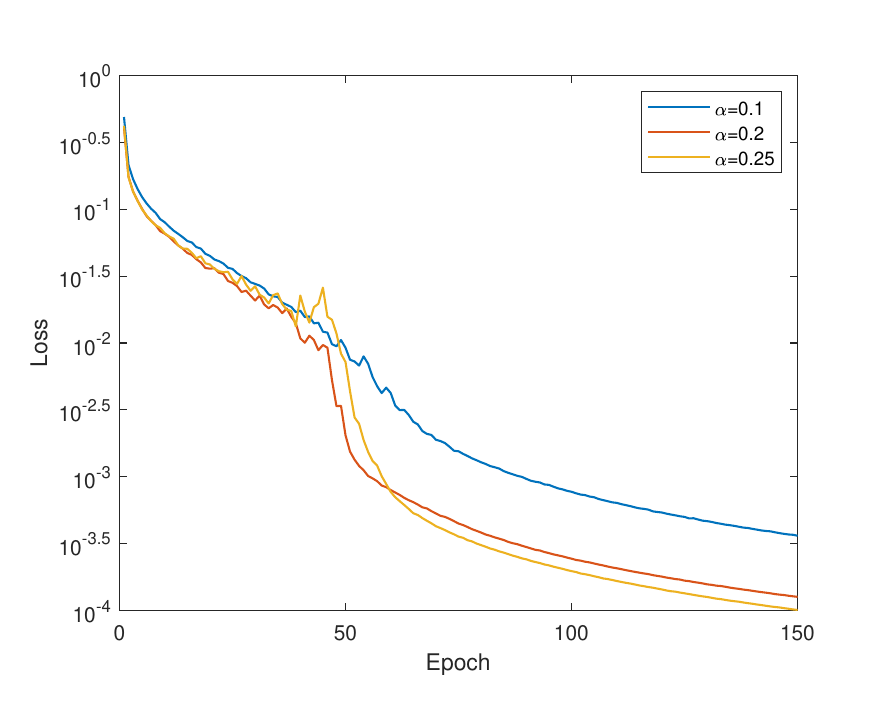} &
\includegraphics[width=0.4\textwidth]{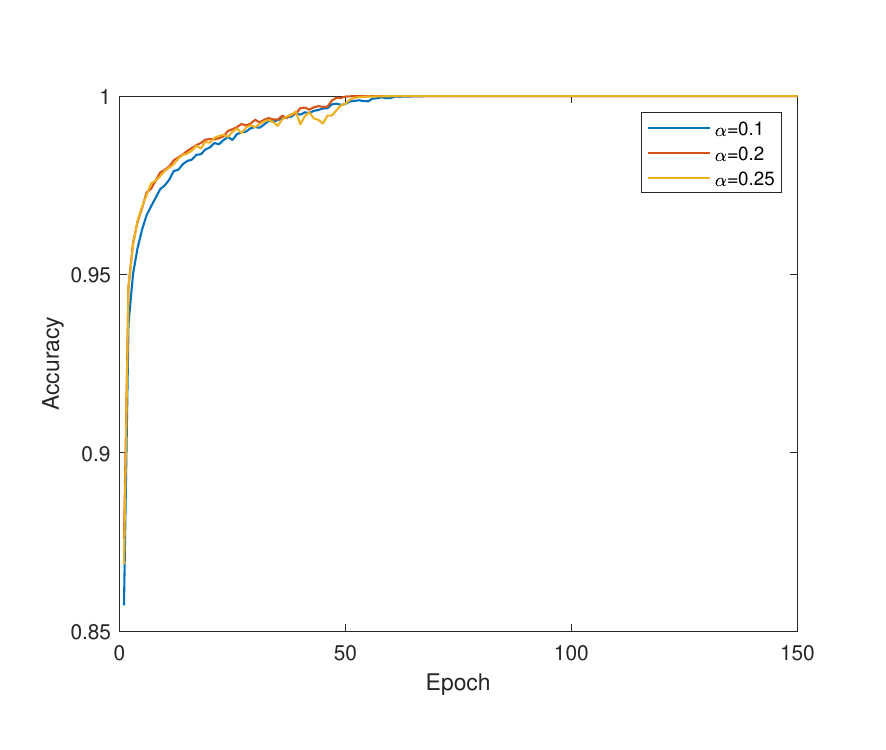} \\
(a) & (b) \\
\end{tabular}
\protect\caption{Plots of loss function and network accuracy during training. (a) As the epoch time increases, the loss function first decreases rapidly. It still descends after $50$ epoch, with much slower rate through, approaching some constant at round $150$ epoch. (b) With the increase of epoch, the network accuracy increases.  By $150$ epoch, the accuracy reaches some ideal value close to $100\%$.
}
\label{LossandAccuracy}
\end{figure*}

Each sample is a mono-color image of $28\times 28 = 784$ pixels, representing the $0$ - $9$ digits. Therefore the input and output layers have $784$ and $10$ nodes respectively. For a typical setting of a fully connected neural network, we add two hidden layers of $30$ neurons each. The mini-batch size is set to $200$, and the network is trained for $150$ epochs in the initial stage (One epoch equals exhausting  through the entire training set of samples once). After the initial phase, the accuracy of the network usually reaches well above $99\%$ (close to $100\%$).  The magnitude of the loss function becomes very small and varies slowly, as shown in Fig.~\ref{LossandAccuracy}. At this point we regard the network as having reached a steady stage. The network parameters are saved for subsequent calls.

\section{Sampling and Analyses of the Neural Weights}

Re-load the trained model and parameters, perform $1$ epoch training, then extract the weight matrix based on this epoch. On a $784\times30\times30\times10$ network, we extract the weights between the two hidden layers, which is a $30\times30$ matrix. It is then stacked into a one-dimensional $900\times1$ weight vector.

\begin{figure}[!ht]
 \centering
 \includegraphics[width=0.8\textwidth]{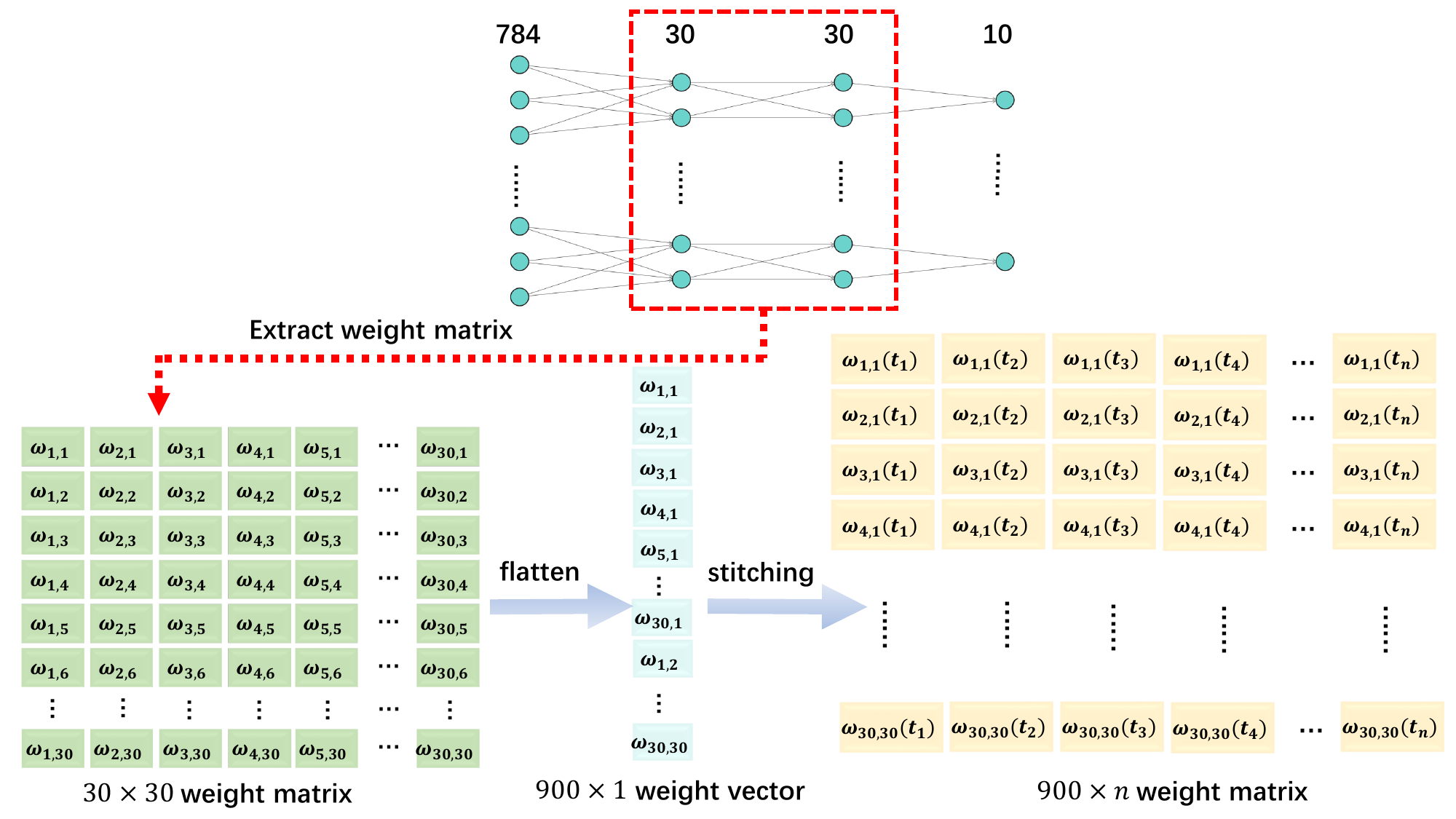}
 \caption{Extraction of weight matrix. The network input layer has 784 neurons, both hidden layers have 30, and the output layer is set to 10 with no bias parameters.}
 \label{ExtractWeightMatrix}
\end{figure}

In SGD, an iteration is performed after each minibatch. There are $n=(N/$batchsize) steps ($N$ being the total number of samples) in one epoch, forming a time window with time series $(t_{1},t_{2},\cdots, t_{n})$. We horizontally stitch the weight vectors of the time series together. After $1$ epoch, we then have a two-dimensional weight matrix with columns of the matrix coming from different batches. For batchsize = $200$, we have $n=300$, therefore a data matrix of $900\times 300$. This is shown schematically in Fig.~\ref{ExtractWeightMatrix}. Further analyses are based on such set of weight matrices.

\begin{figure}[!ht]
 \centering
 \includegraphics[width=0.8\textwidth]{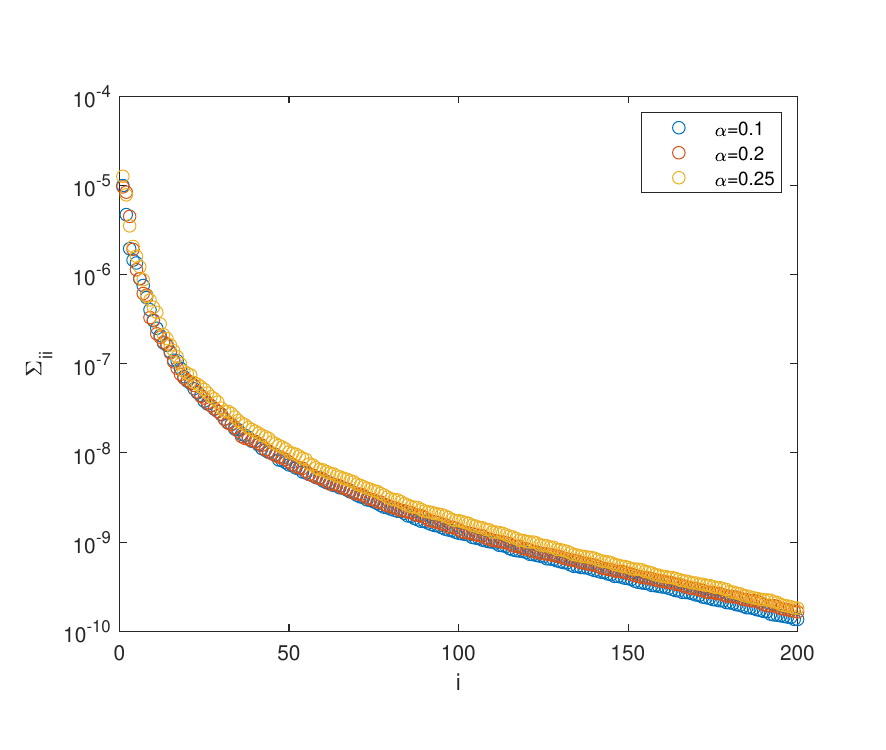}
 \caption{In a fully connected network containing two hidden layers with batchsize = 200, the eigenvalues $\Sigma_{ii}$ of the covariance matrix $\bm{\Sigma}$ for different learning rates $\alpha$. They are virtually identical in these cases.}
 \label{figure_Cii}
\end{figure}

Covariance matrix on the one-dimensional weights with respect to the time series can be obtained by their mean-square fluctuations, cf. Eq.~(\ref{Covariance_matrix}). Singular-value decomposition \cite{strang1993introduction} is used to diagonalize the resulting matrix and arrange the eigenvectors in descending order of the eigenvalues. Some covariance matrices of same network structure but different learning rates are shown in Fig.~\ref{figure_Cii}. The result indicates that the covariance is insensitive to the learning rate.

%

\section{Diffusion Matrix in SGD}

In the update process of SGD, the variance of the mini-batch gradient is also (proportional to) the variance of the SGD noise.  The relationship holds at least near the vicinity of a fixed point. It is a constant but highly anisotropic matrix. In computing the matrix, we adopt the definition from \cite{chaudhari2018stochastic} (cf. below). First we average over the weight matrix of a time series for a set of center-of-mass weight parameters. Import this set of parameters into a trained neural network, overriding the relevant layers already there (The network itself has been trained for $150$ epochs and saved in advance).

These parameters are used as the initial parameters. On this basis, we run the training of the neural network once for 1 epoch again. The purpose of training this time is not to update the parameters, but to derive the gradient of the average weights in order to calculate the diffusion matrix $\vec{D}$. Therefore the optimizer.step (of the underlying python routine) is removed to ensure the weights are kept at the fixed moment. During this training, set the batchsize to $1$ to obtain the gradient corresponding to each of the $60,000$ samples. In the process we use the same splicing method as for the extraction of the weight matrix mentioned above. Finally, Eq.~(\ref{D_def}) of the main work \cite{mainwork} is employed to compute  the desired diffusion matrix with some matrix multiplications.

\section{Reconstruction of $\vec{F}$ Matrix}

Having obtained the $\vec{D}$ matrix, we can put it under the the eigenvectors of the $\bm{\Sigma}$ matrix and let each matrix element be $D_{ij}$.  From Eq.~(\ref{CovarianceMatrixWithForceMatrix}) of the main work \cite{mainwork}, the matrix elements of the $\vec{F}$ matrix can be expressed as:
\begin{equation}
F_{ij}=\frac{2\epsilon D_{ij}}{\Sigma_{ii}+\Sigma_{jj}}.
\end{equation}
Note that $\epsilon \equiv \alpha/2B$ and $\vec{F}$ is symmetric since it comes from the second derivatives of the loss function. The diagonal matrix elements $\vec{F}$, $\vec{D}$ and and $\bm{\Sigma}$, as shown in Fig.~\ref{FDCmatrixForCpre1} below, and the same is replicated as Fig.~\ref{FDCmatrixForCpre} of the main work \cite{mainwork}.
\begin{figure*}[!ht]
\begin{tabular}{ccc}
\includegraphics[width=0.4\textwidth]{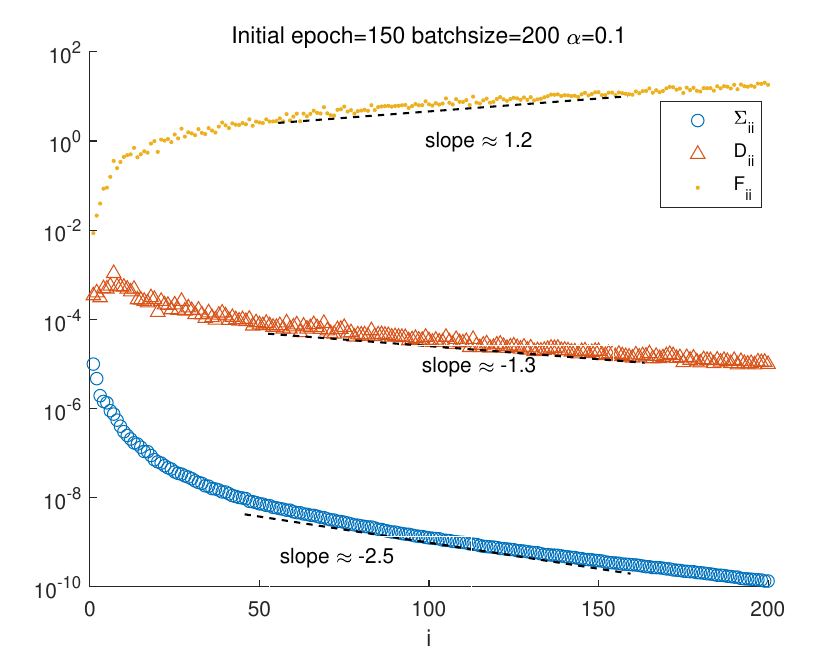} &
\includegraphics[width=0.4\textwidth]{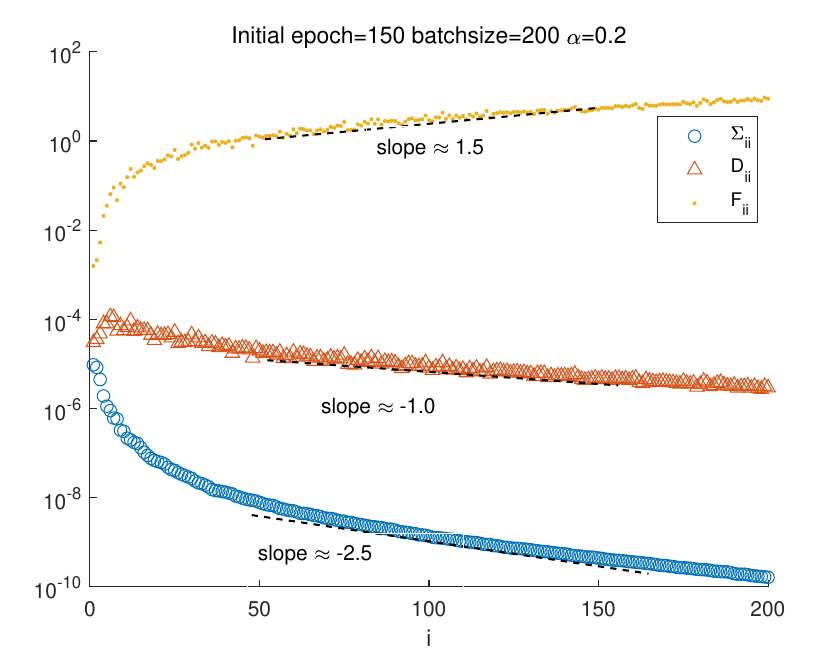} \\
(a) & (b) \\
\end{tabular}
\protect\caption{The relationship between the three matrices at different learning rates under the eigen space of $\bm{\Sigma}$. Here $i$ represents the direction of PCA, starting from the larger eigenvalue of $\Sigma_{ii}$. (a) $\alpha$=1, $\Sigma_{ii}\propto i^{-2.5}$, $D_{ii}\propto i^{-1.3}$, $F_{ii}\propto i^{1.2}$. (b) $\alpha=0.2$, $\Sigma_{ii}\propto i^{-2.5}$, $D_{ii}\propto i^{-1.0}$, $F_{ii}\propto i^{1.5}$.
}
\label{FDCmatrixForCpre1}
\end{figure*}


On a typical set of simulations, we found that for $\alpha=0.1$, $\Sigma_{ii}\propto i^{-2.5}$, $F_{ii}\propto i^{1.2}$; for $\alpha=0.2$, $\Sigma_{ii}\propto i^{-2.5}$, $F_{ii}\propto i^{1.5}$. The key point to observe which is common among simulations of different setup of networks is as follows: $\vec{D}$ also decreases with the PCA index. As a result, we would not have $F_{ii}\times \Sigma_{ii} \sim$ approaching constant under any circumstance, an clear indication of anomaly between variance and flatness were the lost function regarded as the energy function in the SGD process. But such a paradox can be readily resolved when the true energy function is identified as appearing in Eq.~(\ref{Stochastic_Potential}) and Eq.~(\ref{Stationary_distribution}) of the main work \cite{mainwork}, with $\vec{U} = \epsilon\bm{\Sigma}^{-1}$.

In addition,  different learning rates lead to different scaling behavior of $\vec{F}$, which can be visualized from Fig.~\ref{FDCmatrixForCpre}. It occurs despite that $\vec{F}$ does not explicitly involve $\alpha$ as can be seen from Eq.~(\ref{DNN_SDE}). Let us take a snapshot of the matrix elements for comparison. At $\alpha=0.1, i=10$: $F_{ii}\approx10^{-0.4}$, $D_{ii}\approx10^{-3.3}$, $\Sigma_{ii}\approx10^{-6.6}$; But for $\alpha=0.2, i=10$: $F_{ii}\approx10^{-1.0}$, $D_{ii}\approx10^{-4.2}$, $\Sigma_{ii}\approx10^{-6.5}$. Though the covariance for different $\alpha$ basically does not change, both $\textbf{D}$ and hence $\textbf{F}$ do significantly. The larger the value of $\alpha$ is, the smaller the value of $\vec{D}$ is, which in turn bring down the magnitude of $\vec{F}$. It indicates that different learning rates lead the neural network to different parameter spaces.

\section{The Transverse Matrix $\vec{Q}$}

We now evaluate the transverse matrix in the stochastic decomposition as appears in Eq.~(\ref{Force_matrix}) of the main work \cite{mainwork}. It can be obtained via Eq.~(\ref{F_D_U}) in a similar way as for $\vec{F}$. It is unique to our formalism of stochastic dynamics and the knowledge $\vec{Q}$ is a prerequisite to unlock the vortex-like circulations around fixed points found in a related study \cite{PNSwork}. This in turn constitutes the key part for the proposed strategy targeting the well-known catastrophic weight-forgetting difficulty, cf. the main work \cite{mainwork} for further discussions.

\begin{figure*}[!ht]
\begin{tabular}{ccc}
\includegraphics[width=0.4\textwidth]{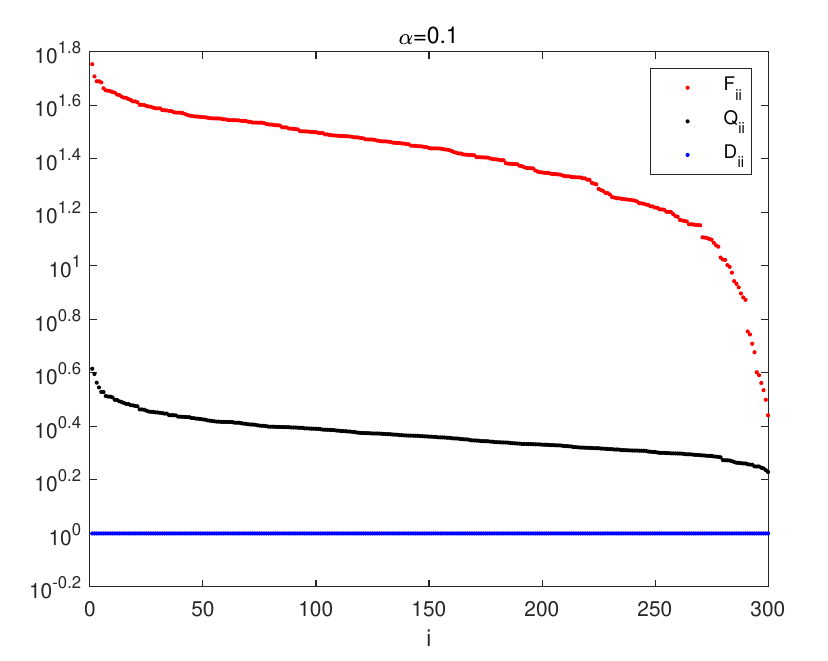} &
\includegraphics[width=0.4\textwidth]{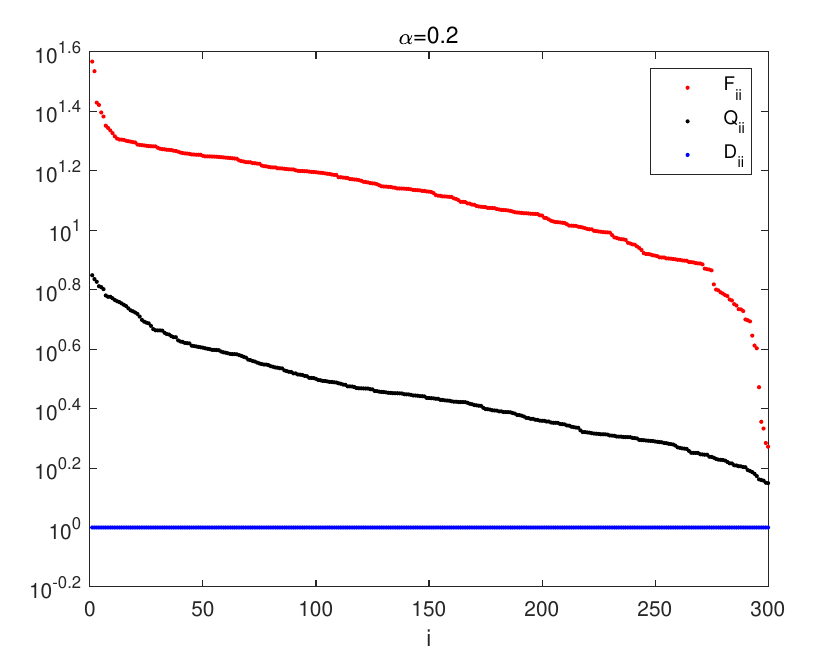} \\
(a) & (b) \\
\end{tabular}
\protect\caption{The relationship between $D_{ii}$, $\tilde{Q}_{ii}$ and $F_{ii}$, when the first is transformed into an identity matrix.(a) $\alpha=0.1$, $D_{ii}: 10^{0}$, $F_{ii}: 10^{0.4}\sim10^{1.8}$, $\tilde{Q}_{ii}: 10^{0.2}\sim10^{0.6}$. (b) $\alpha=0.2$, $D_{ii}: 10^{0}$, $F_{ii}: 10^{0.3}\sim10^{1.6}$, $\tilde{Q}_{ii}: 10^{0.1}\sim10^{0.8}$.
}
\label{QcomputeByuniD}
\end{figure*}

To proceed further, it is most convenient to work under the eigen spaces of the $\vec{F}$ matrix where $\vec{F}$ is diagonal, which can be obtained via singular decomposition as in the case for $\bm{\Sigma}$. We have
\begin{equation}
Q_{ij}= D_{ij}\left(\frac{F_{ii} - F_{jj}}{F_{ii}+F_{jj}}\right).
\end{equation}
As $\vec{Q}$ is anti-symmetric, we can evaluate the diagonal values of the square root of $\vec{Q}\vec{Q}$ instead, denote them by $\tilde{Q}_{ii}$. Some typical results are plotted in Fig.~\ref{QcomputeByuniD}.

\begin{figure*}[!ht]
\begin{tabular}{ccc}
\includegraphics[width=0.4\textwidth]{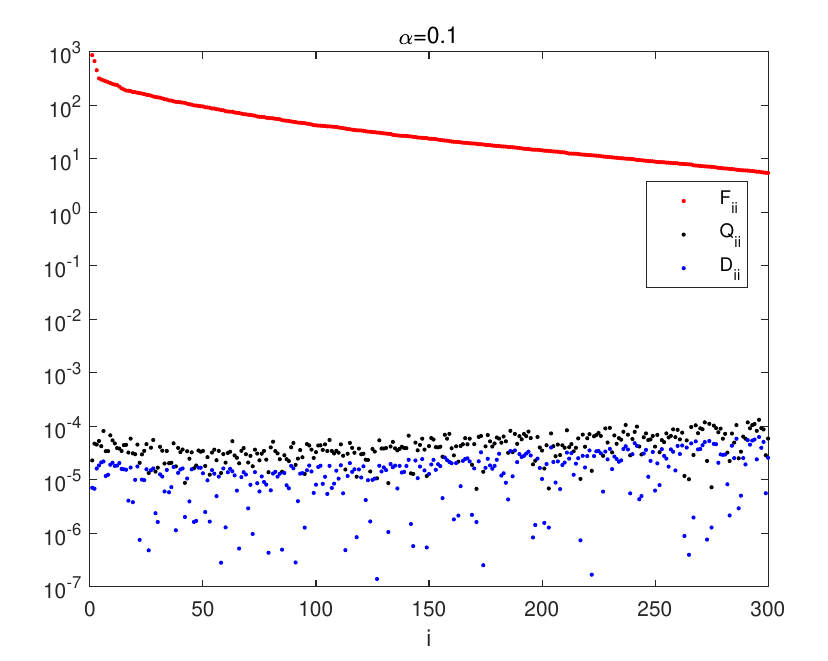} &
\includegraphics[width=0.4\textwidth]{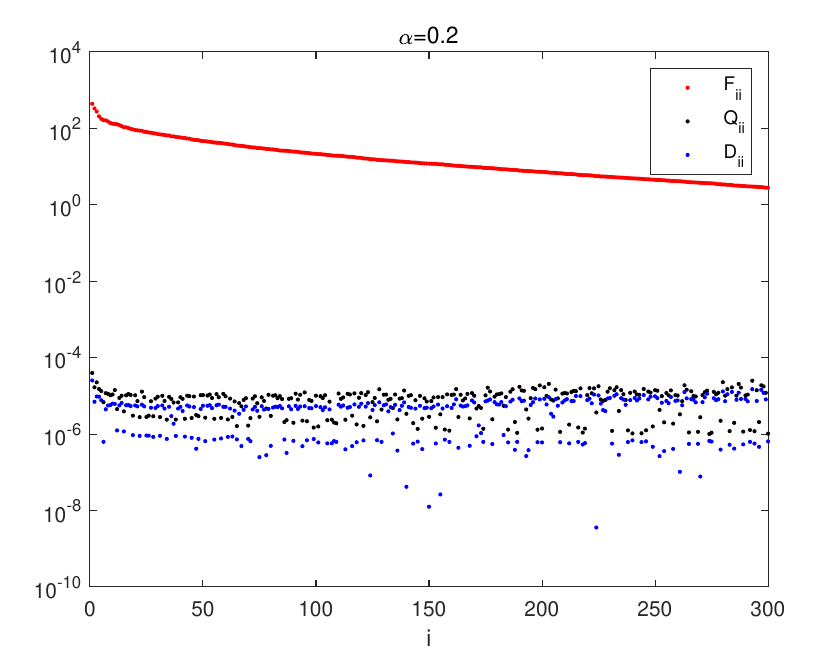} \\
(a) & (b) \\
\end{tabular}
\protect\caption{The relationship between $D_{ii}$, $\tilde{Q}_{ii}$ and $F_{ii}$, (a) $\alpha=0.1$, $D_{ii}: 10^{-6.9} \sim 10^{-4.2}$, $F_{ii}: 10^{0.7}\sim10^{2.9}$, $\tilde{Q}_{ii}: 10^{-5.2}\sim10^{-3.9}$. (b) $\alpha=0.2$, $D_{ii}: 10^{-8.4} \sim 10^{-4.6}$, $F_{ii}: 10^{0.4}\sim10^{2.6}$, $\tilde{Q}_{ii}: 10^{-6.0}\sim10^{-4.4}$.}
\label{QcomputeByNormalD}
\end{figure*}

The result appears to suggest that $\vec{Q}$ is insignificant compared to $\vec{F}$. However, the presence of $\vec{Q}$ is due to the anisotropy in $\vec{D}$. Therefore the correct comparison should be drawn between $\vec{Q}$ and $\vec{D}$ as the two appear in same footing in Eq.~(\ref{Force_matrix}) of the main work \cite{mainwork}. To see this we can re-scale the state variables such that $\vec{D}$ becomes the unit matrix $\vec{I}$.  A general (non-orthogonal) coordinate transformation\cite{Kwon13029} $\vec{x}\rightarrow \vec{y}=\vec{M}^{-1}\vec{x}$, changes
\begin{equation}
\textbf{F}\rightarrow \textbf{M}^{-1}\textbf{F}\textbf{M},
\label{TransformF}
\end{equation}
\begin{equation}
\textbf{U}\rightarrow \textbf{M}^{\tau}\textbf{U}\textbf{M}.
\label{TransformD}
\end{equation}\begin{equation}
\textbf{D}\rightarrow \textbf{M}^{-1}\textbf{D}(\textbf{M}^{\tau})^{-1}.
\label{TransformD}
\end{equation}
And $\vec{Q}$ transforms the same way as $\vec{D}$.  The superscript $\tau$ in the above denotes the transverse of the underlying matrix. Note that $\vec{F}$ is no longer symmetric under the transformation. 
But the results are drastically different, as shown in Fig.~\ref{QcomputeByuniD}.


It is evident that once the system is re-scaled to have an isotropic diffusion process.  $\vec{Q}$ is larger compared to $\vec{D}$, offering an ideal environment that leads to vortex circulations as seen in \cite{PNSwork}. Such characteristic plays a pivot role for the algorithm improvement discussed at the end of the main work \cite{mainwork}.

\section{Data Availability}
The necessary formulae and the steps to perform the computations are detailed in the main work and the Supplementary Information. The version of python used in this work is 3.9.5, and the version of torch is 1.9.0. For the construction of neural network refer to the contents of an online blog \cite{NeuralNetworks}. The data used to justify the results and conclusions of this work are entirely presented within the body and supplementary information of the manuscript.

The code used in this paper is available on request from chenyongcong@shu.edu.cn.

%